\documentclass[12pt,amsmath,amssymb]{article}
\usepackage[usenames]{color}
\usepackage{graphicx}

\begin{document}

\thispagestyle{empty}
\def\beq{\begin{eqnarray}}
\def\eeq{\end{eqnarray}}   

\noindent\textbf{\textcolor{black}{Possible polarisation and spin dependent  aspects} }\\
\textcolor{black}{\textbf{of quantum gravity}}
\vskip 0.75cm

\noindent\textit{D.~V.~Ahluwalia-Khalilova, N.~G.~Gresnigt, Alex~B.~Nielsen,\\ 
D.~Schritt, T.~F.~Watson}
\vskip 0.5cm

\noindent
{\footnotesize Department of Physics and Astronomy, Rutherford Building\\
University of Canterbury\\
Private Bag 4800, Christchurch 8020, New Zealand\\
E-mail: dharamvir.ahluwalia-khalilova@canterbury.ac.nz\\
}

\vskip 0.5cm
\textcolor{red}{\hrule}

\vskip .3cm


\noindent
\textbf{\small We argue that
quantum gravity theories that carry a Lie algebraic modification
of the Poincar\'e and Heisenberg algebras
inevitably provide
inhomogeneities that may serve as seeds for cosmological 
structure formation.
Furthermore, in this class of theories one must
expect a strong
 polarisation and spin dependence of various
quantum-gravity effects.}
\vskip 0.75cm

\noindent
\textit{I. Introduction\textemdash~}
Quantum gravity proposals often come with
a modification of the Heisenberg, and Poincar\'e, algebras. Confining 
ourselves to
Lie algebraic modifications, 
we argue that the underlying physical space of all 
such theories must be inhomogeneous. In order to establish
this result, we first  
review how, within
a quantum framework, the homogeneity  and continuity of physical space  
 lead inevitably to
the Heisenberg algebra. 
We then review general arguments 
that hint towards
 algebraic modifications encountered in quantum gravity proposals.
Next,
we argue that a natural extension of physical laws to the Planck
scale can be obtained by a Lie algebraic modification of 
the Poincar\'e and Heisenberg algebras in such a way that the 
resulting algebra is \textit{immune} to infinitesimal perturbations 
in its structure constants.
With the  context so chosen, we
 establish the main thesis: that
quantum gravity theories of the aforementioned  class
inevitably provide
inhomogeneities that may serve as seeds for structure formation; and
that  quantum gravity induced effects may carry a strong
 polarisation and spin dependence.

The established results are
not restricted to the chosen algebra but may easily be extended to
all Lie algebraic modifications that alter the Heisenberg algebra.\footnote{A 
slightly weaker argument
can be constructed for non-Lie algebraic proposals when we confine
ourselves to  probed distances significantly larger than the
length scale associated with the loss of spatial continuity.}
\vskip 0.54cm

\noindent
\textit{2. Homogeneity and continuity of physical space, and its
imprint in the
Heisenberg algebra\textemdash~}
In order to understand the fundamental origin of primordial
inhomogeneities we will first review the
fundamental connection between the homogeneity and continuity
of physical space and the Heisenberg algebra.  It is in this spirit
that we remind our reader of an argument that is presented, for example,
by Isham in~\cite[Section 7.2.2]{Isham:1995ci}. 
There it is shown that, in  the general quantum mechanical framework, 
and under the following two assumptions,
\begin{enumerate}
\item[\textemdash~] physical space is homogeneous, 
\item[\textemdash~] any spatial distance $r$ can be divided
in to two equal parts, $r=r/2+r/2$,
\end{enumerate}
it necessarily follows
that the operator ${x}$ associated with  position measurements
along the $x$-axis, and the 
 generator of displacements ${d}_x$ 
along the $x$-direction, satisfy
$
[{x},{d}_x]= i.
$
If one now requires consistency with the elementary wave mechanics 
of Heisenberg,
one must identify ${d_x}$ 
with ${p}_x/\hbar$ (${p}_x$ is the operator associated with
momentum measurements along the
$x$-direction). This gives,
$
\left[{x},{p}_x\right] = i \hbar.
$
Without any additional assumptions, the argument easily
generalises to yield the entire Heisenberg algebra
$
\left[{x}_j,{p}_k\right] = i 
\hbar \delta_{j k},\quad \left[{p}_j,
{p}_k\right] = 0,\quad \left[{x}_j,{x}_k\right] = 0,
$
where ${x}_j$, $j=1,2,3$, are the position operators associated with
the three coordinate axes, where  the observer is assumed to be located
at the origin of the coordinate system.

Thus it is evident that
a quantum description
of physical reality,
with 
  spatial \textit{homogeneity} and \textit{continuity},
inevitably leads to the    
Heisenberg algebra.
\vskip 0.54cm

\noindent
\textit{3. On the need to go beyond the Heisenberg and Poincar\'e
algebraic-based description of physical reality\textemdash~}
From an algebraic point of view 
much of the success of modern physics can be traced back
to the Poincar\'e and Heisenberg algebras.
Had the latter algebra been discovered before the former,
the conceptual formulation and evolution of theoretical 
physics would have been significantly different. For instance,
it is a direct implication of Heisenberg's fundamental commutator
$[{x}_i, {p}_j] = 
i \hbar \delta_{ij} \; (\mbox{with}~ i,j=1,2,3)$, 
that \textit{events} should be characterised 
not only by their spatiotemporal location
$x_\mu$, but also by
the  associated
energy momentum $p_\mu$;
and  that should be done in a manner consistent with
the fundamental measurement uncertainties inherent in the
formalism. 
The reader may wish to come back to these remarks 
in the context of Eq.~(\ref{eq:px}) where one
shall find that in a specific  sense the  physical
space that underlies the conformal algebra does indeed
combine the notions of spacetime and energy momentum. 
Furthermore, as will be seen from Eq.~(\ref{eq:18}) and the subsequent
remarks, this
interplay becomes increasingly important as we consider the
early universe above  $\approx 100\;\mbox{GeV}$.

 In the mentioned  
description the interplay of the general relativistic and quantum
mechanical frameworks becomes inseparably bound.
To see this, consider the well-known  thought experiment
to probe spacetime at spatial resolutions around  
the
Planck 
length $\ell_P \stackrel{\mathrm{def}}{=}\sqrt{\hbar G/c^3}$. If
one does that, one ends up creating
a Planck mass $m_P  \stackrel{\mathrm{def}}{=} \sqrt{\hbar c/G}$ black hole. 
This fleeting 
structure carries a temperature $T \approx 10^{30} K$ and evaporates
in a thermal explosion in $\approx 10^{-40}$ seconds. This, incidentally, 
is a long time \textendash~ about ten thousand fold the Planck time $\tau_P  
\stackrel{\mathrm{def}}
{=}\sqrt{\hbar G/ c^5}$.
The formation and evaporation of the black hole 
places a fundamental limit on the spatiotemporal
resolution with which spacetime can be probed. 

The authors of \cite{Ahluwalia:1993dd,Doplicher:1994zv}
have argued that once gravitational effects associated with
the quantum measurement process are accounted for, the Heisenberg
algebra, and in particular the commutator 
$\left[{x}_j,{p}_k\right]$, must be modified. The role of gravity in 
the quantum measurement process was 
also emphasised by Penrose~\cite{Penrose:1996cv}.

\begin{quote}
\textit{From the above discussion, we take it as suggestive that
an operationally-defined view of physical space (or, its generalisation)
shall inevitably ask for  the length scale, $\ell_P$
to play an important role.} 
\end{quote}
In the context of the continuity of
physical space we will take it as a working hypothesis that, just
as a lack of commutativity of the $x$ and $p_x$ operators does not render
the associated eigenvalues discrete, similarly the existence of 
a non-vanishing $\ell_P$ does not necessarily make the underlying
space lose its continuum nature. This
is a highly  non-trivial issue 
requiring a detailed discussion from which 
we here refrain;  yet, an element of justification 
shall become apparent below.

From a dynamical point of view, as early as late 1800's,
the symmetries of Maxwell's equations were
already suggesting a  merger of space and time 
into one physical entity, spacetime \cite{Brown:2005}. Algebraically, 
these symmetries are encoded in the Poincar\'e algebra. The emergent 
unification of space and time called for a new fundamental invariant, $c$,
the speed of light (already contained in Maxwell's equations). From an
empirical point of view, the Michelson-Morley experiment 
established the constancy of the speed of light for all
inertial observers, and thus re-confirmed, in the Einsteinian framework,
the implications of the Poincar\'e spacetime symmetries.

Concurrently, we note that while in classical statistical mechanics
it is the volume that determines the number of accessible states and hence
the entropy, the
situation is dramatically different in a gravito-quantum mechanical setting.
One example of this assertion may be found in 
the well-known Bekenstein-Hawking entropy result for a 
Schwarzschild black hole,
$
S_{BH} = (k/4)(A/\ell_P^2)
$;
where $k$ is the Boltzmann constant, and $A$ is the surface area of the 
sphere contained within the event horizon of the black hole. Thus quantum
mechanical and gravitational realms conspire to suggest the holographic
conjecture \cite{'tHooft:1993gx,Susskind:1994vu,deHaro:2001}. 
The underlying physics is perhaps two fold: (a) contributions
from higher momenta in quantum fields to the number of accessible states 
is dramatically reduced because these are screened by the associated event 
horizons; and (b) the accessible states for a quantum system are
severely influenced by the behaviour of the wave function
at the boundary.

\begin{quote}
\noindent
\textit{From this discussion, we take it as suggestive that
in quantum cosmology/gravity the new operationally-defined 
view of physical space shall inevitably ask for a cosmological 
length scale, 
$\ell_C$.}
\end{quote}
These observations prepare us to reach the next 
trail in our essay.

In the immediate aftermath of cosmic creation with the big bang, the
physical reality knew of no inertial frames of Einstein. This is due
to the fact that massive particles had yet to appear on the scene. The
spacetime symmetries at cosmic creation are encoded in the conformal
algebra.  So, whatever new operational view of spacetime emerges, it
must somehow also incorporate a process by which one evolves from the
``conformal phase'' of the universe at cosmic creation to the present
(see Fig.~1).

\begin{quote}\noindent
\textit{Algebraically, we take it to suggest that there must be
a mechanism that describes how the present day Poincar\'e-algebraic
description relates to the conformal-algebraic description of the 
universe at its birth.}
\end{quote}
 We parenthetically
note that in the conformal phase, where leptons and quarks 
were yet to acquire mass (through the Higg's mechanism, or something of
that nature), the operationally-accessible symmetries are not
Poincar\'e but conformal.
This is so because 
to define rest frames, so essential for
operationally establishing the Poincar\'e algebra,
one needs massive particles.
In the transition when massive particles come to exist, 
the local algebraic
symmetries of general relativity
suffer an operational change. Consequently, for the cosmic
epoch before $\approx 100\;\mbox{GeV}$ general relativistic description
of physical reality might require modification.   
 
\vskip 0.54cm

\noindent
\textit{4. A new algebra for quantum gravity and the emergent
 inhomogeneity of physical space\textemdash~}
Mathematically, a Lie algebra incorporating the three italicised items
in Sec.~3 already exists. It was inspired by Faddeev's
mathematical analysis of the quantum and relativistic revolutions of
the last century~\cite{Faddeev:1989} and was followed up by Vilela
Mendes in his 1994 paper~\cite{VilelaMendes:1994zg}.  The uniqueness
of the said algebra was then explored through a Lie-algebraic
investigation of its stability by Chryssomalakos and Okon,
in 2004 \cite{Chryssomalakos:2004gk}.  Some of the physical
implications were subsequently explored in
Refs.~\cite{Ahluwalia-Khalilova:2005jn,Ahluwalia-Khalilova:2005km},
and its Clifford-algebraic representation was provided by Gresnigt
\textit{et al.} \cite{Gresnigt:2006gq}.
Its importance was further noted in \textit{CERN Courier} \cite{cern:2005}.

However, its candidacy for the algebra underlying quantum
cosmology/gravity has been difficult to assert. This is essentially 
due to a
perplexing observation made in Ref.~\cite{Chryssomalakos:2004gk}
regarding the interpretation of the operators associated 
with the spacetime events. In this essay we overcome this 
interpretational hurdle and argue that it contains all the desired
features for such an algebra.

To this end we first write down what has come to be known
as the Stabilised Poincar\'e-Heisenberg Algebra (SPHA) and then proceed
with the interpretational issues.  
The SPHA contains the Lorentz sector 
(we follow the widespread physics convention 
which takes the $\mathcal{J}_{\mu\nu}$ as dimensionless and  $ \mathcal{P}_\nu$
as dimensionful)
\beq
 \left[\mathcal{J}_{\mu\nu},\mathcal{J}_{\rho\sigma}\right] = 
i \left(
 \eta_{\nu\rho} \mathcal{J}_{\mu\sigma}+\eta_{\mu\sigma} \mathcal{J}_{\nu\rho}  
 - \eta_{\mu\rho} \mathcal{J}_{\nu\sigma}  
- \eta_{\nu\sigma} \mathcal{J}_{\mu\rho} \right)
\eeq
This remains unchanged (as is strongly suggested by the
analysis presented in~\cite{Collins:2004bp}), as does the commutator
\beq
 \left[\mathcal{J}_{\mu\nu}, \mathcal{P}_\lambda\right] 
= i \left(\eta_{\nu\lambda} \mathcal{P}_\mu 
- \eta_{\mu\lambda} \mathcal{P}_\nu \right)
\eeq
These are supplemented by the following modified sector
\beq
&& \left[\mathcal{J}_{\mu\nu}, \mathcal{X}_\lambda\right] = 
i \left(\eta_{\nu\lambda} \mathcal{X}_\mu 
- \eta_{\mu\lambda} \mathcal{X}_\nu \right)\\ 
&& \left[\mathcal{P}_\mu,\mathcal{P}_\nu\right] = 
 i q \alpha_1 \mathcal{J}_{\mu\nu}\\
&& \left[\mathcal{X}_\mu,\mathcal{X}_\nu\right] =  i q \alpha_2 
\mathcal{J}_{\mu\nu} \label{eq:zncst}\\
&& \left[\mathcal{P}_\mu,\mathcal{X}_\nu\right] = 
i  q \eta_{\mu\nu} {\mathcal{I}}  + i q \alpha_3\, 
\mathcal{J}_{\mu\nu}\label{eq:zhfc}\\ 
&& \left[\mathcal{P}_\mu,\mathcal{I}\right] =
 i \alpha_1 {\mathcal{X}_{\mu}} - i \alpha_3 \mathcal{P}_\mu \\
&& \left[\mathcal{X}_\mu,{\mathcal{I}}\right] = i \alpha_3 {\mathcal{X}}_\mu 
- i \alpha_2 \mathcal{P}_{\mu}
\\
&& \left[\mathcal{J}_{\mu\nu},\mathcal{I}\right] = 0\label{eq:za2}
\eeq 
The metric $\eta_{\mu\nu}$ is taken to have the signature
$(1,-1,-1,-1)$. The SPHA is stable, except 
for the instability surface defined by $\alpha_3^2 = \alpha_1 \alpha_2$ (see
Fig.~2).
Away from the instability surface the SPHA is immune to infinitesimal
perturbations in its structure constants. This distinguishes SPHA from
many of the competing algebraic structures because  a physical theory 
based on such an algebra
is likely to be free from ``fine tuning'' problems. This is essentially self
evident
because if an algebraic structure does not carry this immunity,
one can hardly expect the physical theory based upon  such an
algebra to enjoy the opposite.

The SPHA involves three parameters $\alpha_1,\alpha_2,\alpha_3$. 
The $c$ and $\hbar$  arise in the process of the 
Lie algebraic stabilisation that  takes
us from the Galilean relativity to Einsteinian relativity, and 
from classical mechanics to quantum mechanics.
Their specific values are fixed by experiment. Similarly, 
 $\alpha_1,\alpha_2,\alpha_3$ owe their origin to a similar stabilisation
of the \textit{combined} Poincar\'e and Heisenberg algebra.

Except for the fact that $\alpha_1$ must be 
a measure of the size of the observable  universe (here 
assumed  to be operationally determined from the Hubble parameter),
the Lie algebraic procedure for obtaining SPHA does not determine
$\alpha_1,\alpha_2,\alpha_3$.  Dimensional and phenomenological 
considerations, along with the requirement that we  
obtain physically viable limits,
suggest
the following identifications:\footnote{In making
the identifications it is understood that these may be true up to
a multiplicative factor of the order of unity.}
\beq
\alpha_1 := \frac{\hbar}{\ell_C^2} \label{eq:alpha1}
\eeq
where $\ell_C$ is of the order of the Hubble radius,
 and therefore  
it depends on the cosmic epoch.
The introductory remarks, and existing data suggest that~\cite{Chryssomalakos:2004gk}
\beq
\alpha_2 = \frac{\ell_P^2}{\hbar}   \label{eq:alpha2} 
\eeq

In the limit $\ell_P\to 0,\ell_C\to \infty, \beta
\to 0, \mathcal{I}\to I, \mbox{the identity operator}$, 
the SPHA splits into 
 Heisenberg and Poincar\'e
algebras.
In that limit, the symbols
${\mathcal{X}_\mu}\to {x}_\mu,{\mathcal{P}_\mu} \to {p}_\mu, 
\mathcal{J}_{\mu\nu}\to J_{\mu\nu}$, and ${\mathcal{I}}\to I$.
Thus  ${x}_\mu, {p}_\mu, J_{\mu\nu}, I$
acquire their traditional
meaning, while
$\mathcal{X}_\mu,\mathcal{P}_\mu, 
\mathcal{J}_{\mu\nu}, \mathcal{I}$ are to be considered their generalisations.
In particular $x_\mu$  
should then be interpreted as the 
generator of energy-momentum 
translation. The latter
parallels the canonical interpretation of $p_\mu$ as the generator of
spacetime translation.  This interpretation,
we believe, removes the problematic interpretational aspects 
associated with $\mathcal{X}_\mu$ in the analysis of Ref.~\cite{Chryssomalakos:2004gk}.

The identification of $q$ with $\hbar$ is dictated by  the demand
that we recover the
Heisenberg algebra. It also suggests that at the present cosmic epoch 
$\alpha_3$ should not allow the second term in
the right hand side of equation~(\ref{eq:zhfc}) to have a   
significant contribution.
It will become  
apparent below that $\alpha_3$ is intricately connected to the conformal algebraic 
limit of SPHA. 
With these identifications, and with $\alpha_3$ renamed as the dimensionless
parameter $\beta$, the
SPHA takes the form 

\beq
&& \left[
\mathcal{J}_{\mu\nu},\mathcal{J}_{\rho\sigma}\right] = 
i \left(
 \eta_{\nu\rho} \mathcal{J}_{\mu\sigma}+\eta_{\mu\sigma} \mathcal{J}_{\nu\rho}  
 - \eta_{\mu\rho} \mathcal{J}_{\nu\sigma}  
- \eta_{\nu\sigma} \mathcal{J}_{\mu\rho} \right)\label{eq:a}\\
&& \left[\mathcal{J}_{\mu\nu}, \mathcal{P}_\lambda\right] = 
i \left(\eta_{\nu\lambda} \mathcal{P}_\mu 
- \eta_{\mu\lambda} \mathcal{P}_\nu \right),\;
 \left[\mathcal{J}_{\mu\nu}, \mathcal{X}_\lambda\right] = i 
\left(\eta_{\nu\lambda} \mathcal{X}_\mu 
- \eta_{\mu\lambda} \mathcal{X}_\nu \right)\label{eq:b}\\
&& \left[\mathcal{P}_\mu,\mathcal{P}_\nu\right] = 
 i \left({\hbar^2}/{\ell^2_C}\right) {\mathcal{J}_{\mu\nu}},\;
 {\left[{\mathcal{X}_\mu},{\mathcal{X}_\nu}\right] }=  i {\ell^2_P} 
{\mathcal{J}}_{\mu\nu},\;
 \left[{\mathcal{P}}_\mu,{\mathcal{X}}_\nu\right] = 
i  \hbar \eta_{\mu\nu} \mathcal{I}  + i \hbar  \beta\, 
\mathcal{J}_{\mu\nu}\label{eq:hfc} \label{eq:c}\\
&& \left[\mathcal{P}_\mu,\mathcal{I}\right] =
 i \left({\hbar}/{\ell^2_C}\right) {\mathcal{X}}_{\mu} 
- i \beta {\mathcal{P}}_\mu,
\;
 \left[\mathcal{X}_\mu,\mathcal{I}\right] = i \beta \mathcal{X}_\mu 
- i \left({\ell^2_P}/{\hbar}\right) \mathcal{P}_{\mu},\;
\left[\mathcal{J}_{\mu\nu},\mathcal{I}\right] = 0\label{eq:a2}
\eeq

Since cosmic creation began with massless particles, it should be
encouraging if in some limit SPHA reduced to the conformal algebra. 
This is indeed the case. It follows from a 
somewhat lengthy, though simple,  exercise.  
Towards examining this question
we introduce two new operators
\beq
\tilde{\mathcal{P}}_\mu= a  {\mathcal{P_\mu}} + b {\mathcal{X_\mu}},\quad
\tilde{\mathcal{X}}_\mu= a^\prime  {\mathcal{X_\mu}} + b^\prime {\mathcal{P_\mu}} \label{eq:px}
\eeq
and find that if the introduced parameters $a,b,a^\prime,b^\prime$ satisfy the
the following conditions
\beq
a=\frac{\ell_P^2}{b^\prime \hbar},\quad b=\frac{1-\beta}{b^\prime},\quad
a^\prime=\frac{b^\prime \hbar}{\ell_C^2(1-\beta)}  \label{eq:px2}
\eeq
with $\beta^2$ restricted to the value ${1+(\ell_P^2/\ell_C^2)}$,
then SPHA written in terms of 
$\tilde{\mathcal{P}}_\mu$ 
and $\tilde{\mathcal{X}}_\mu$ satisfies the conformal
algebra~\cite[Sec. 4.1]{DiFrancesco:1997ph}.

Using these results, we can re-express
$\tilde{\mathcal{P}}_\mu$ and $\tilde{\mathcal{X}}_\mu$
in a fashion that supports the view taken in the 
opening paragraph of this section 
\beq
\tilde{\mathcal{P}}_\mu = a
\left( {\mathcal{P}}_\mu + \frac{\hbar}{\ell_P^2} 
\left(1-\beta\right) {\mathcal{X}}_\mu\right),
\quad
\tilde{\mathcal{X}}_\mu =  a^\prime
\left( {\mathcal{X}}_\mu + \frac{\ell_C^2}{\hbar}
\left(1-\beta\right)  {\mathcal{P}}_\mu\right) \label{eq:18}
\eeq
with
\beq
\beta^2={1+\frac{\ell_P^2}{\ell_C^2}}.
\eeq
Near the big bang, $\ell_C\approx\ell_P$ 
and thus $\beta\to \pm \sqrt{2}$ (see, Fig.~1). This results in a significant mixing of
the ${\mathcal{X}}_\mu$ and ${\mathcal{P}}_\mu$ in the conformal
algebraic description in terms of 
$\tilde{\mathcal{X}}_\mu$
and $\tilde{\mathcal{P}}_\mu$.

In contrast, hypothetically, had we been on the conformal surface at present 
then taking $\ell_C\gg\ell_P$ makes $\beta\to\pm 1$.
Consequently, for $\beta \to +1$,  
$\tilde{\mathcal{P}}_\mu$ becomes identical 
to ${\mathcal{P}}_\mu$ up to a multiplicative scale factor $a$. 
Similarly, 
$\tilde{\mathcal{X}}_\mu$ becomes identical 
to ${\mathcal{X}}_\mu$ up to a multiplicative scale factor $a^\prime$.
As is evident from Eq.~(\ref{eq:px2}), the multiplicative scale factors
$a$ and $a^\prime$ are constrained by the relation 
$a a^\prime = {\ell_P^2}/({\ell_C^2 (1-\beta)})$. We expect that similar 
modifications to spacetime symmetries would occur if we were to explore
it at Planckian energies in the present epoch.  For $\beta \to -1$ 
( $\ell_C\gg\ell_P$), one again
obtains
 significant mixing of
the ${\mathcal{X}}_\mu$ and ${\mathcal{P}}_\mu$.

By containing $\ell_P$ and $\ell_C$, the SPHA unifies the extreme microscopic
with the extreme macroscopic, i.e., the cosmological. In the early universe it
allows for the existence of conformal symmetry. 
The significant departure from the Heisenberg algebra 
at big bang, yields primordial inhomogeneities in the underlying 
physical space and the quantum fields that it supports. 
The latter is an unavoidable consequence of
the discussion presented in Sec.~2.\footnote{Any one 
of the other suggestions in quantum gravity that modify the
Heisenberg algebra (see, e.g., references \cite{Garay:1994en}-\cite{Ahluwalia:2000iw}) carry similar implications 
for homogeneity and isotropy of the physical space.}
\vskip 0.54cm

\noindent
\textit{5. Polarisation and spin dependence of the cosmic inhomogeneities
and other quantum gravity effects\textemdash~}
A careful examination of SPHA presented in equations~(\ref{eq:a}-\ref{eq:a2})
reveals a strong  $\mathcal{J}_{\mu\nu}$ dependence of the modifications 
to the Heisenberg algebra. Physically,  
this translates to the following representative implications
\begin{enumerate}

\item[\textemdash~]The induced primordial cosmic inhomogeneities are dependent on
spin and polarisation of the fields for which these are calculated.

\item[\textemdash~]The operationally-inferred commutativity/non-commutativity 
of the physical space depends on the spin and polarisation of the probing
particle.

\item[\textemdash~]The just enumerated observation implies that
a violation of equivalence principle is inherent in the SPHA based
quantum gravity.

\item[\textemdash~]Since  Heisenberg algebra  uniquely
determines the nature of the wave particle
duality~\cite{Kempf:1994su,Ahluwalia:2000iw} (including the 
de Broglie result ``$\lambda=h/p$''), it
 would undergo spin and polarisation 
dependent changes in quantum gravity  based on SPHA.

\end{enumerate}
All these results carry over to any theory of quantum gravity
that modifies the Heisenberg algebra with a  $\mathcal{J}_{\mu\nu}$
dependence.
\vskip 0.54cm

\noindent
\textit{6. Conclusion\textemdash~}
In this essay we have motivated a new candidate for the algebra 
which may underlie a physically viable and consistent theory
of quantum cosmology/gravity.
Besides yielding an algebraic unification of the extreme microscopic and cosmological
scales, it generalises the notion of conformal symmetry. 
The modifications to the Heisenberg algebra at the
present cosmic epoch are negligibly small;
but when $\ell_C$ and $\ell_P$ are of
the same order (i.e, at, and near, the big bang), the induced inhomogeneities 
are intrinsic to the nature of physical space. These can then be
amplified by the cosmic evolution and result in 
important back reaction effects
\cite{Buchert:2005xf,Wiltshire:2007jk,Rasanen:2006kp,Ishibashi:2005sj}.
An important aspect of the SPHA-based quantum gravity is that it
inevitably provides
inhomogeneities that may serve as an important ingredient
for structure formation
\cite{Perez:2005gh}.
Furthermore, in this class of theories one must
expect a strong
 polarisation and spin dependence of various
quantum-gravity effects.

\vskip 0.54cm

\noindent
\textit{Acknowledgements\textemdash~We wish to thank Daniel Grumiller and Peter West for their
insightful questions and suggestions.}

\newpage

\newpage
\begin{figure}
\includegraphics{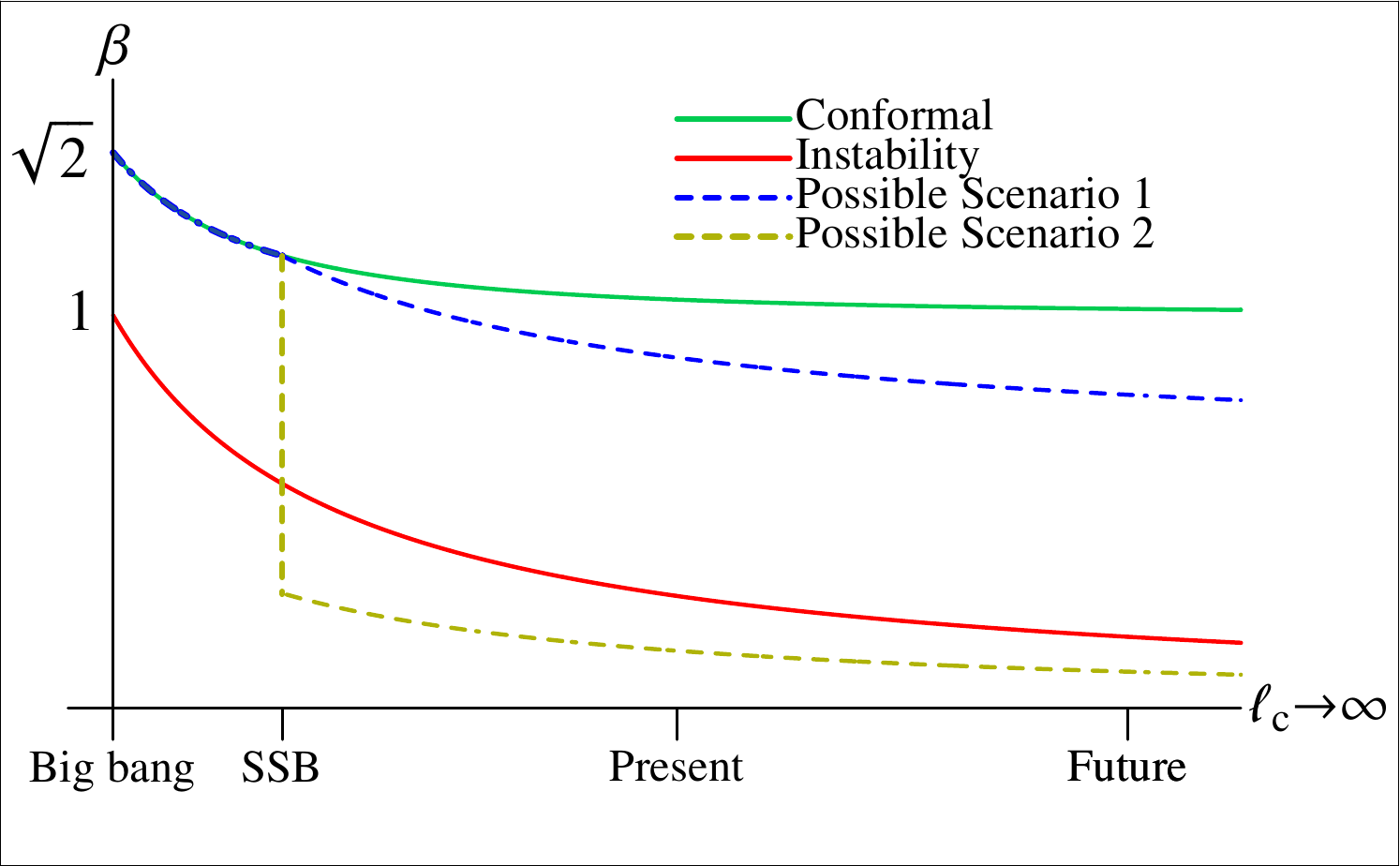}
\caption{
This figure  is a cut, at $\ell_P=1$ (with $\hbar$ set to unity),
of Fig.~2 and it  schematically shows the cosmic evolution along two
possible scenarios. 
For this purpose, only $\beta\ge 0$ values have
been taken. The   $\beta< 0$ sector can easily be inferred from
symmetry consideration.
 In one of the scenarios the 
conformal symmetry  of the early universe is lost without
crossing the instability surface, while in the other it crosses
that surface. In the latter case the algebra 
changes~\cite{Chryssomalakos:2004gk} from
$so(2,4)$ to $so(1,5)$. This crossover, we speculate, 
 may be related to the mass-generating
process of spontaneous symmetry breaking (SSB) of the standard model of
high energy physics. The big bang is here identified with 
$\ell_C \approx \ell_P$.}
\end{figure}

\begin{figure}
\includegraphics{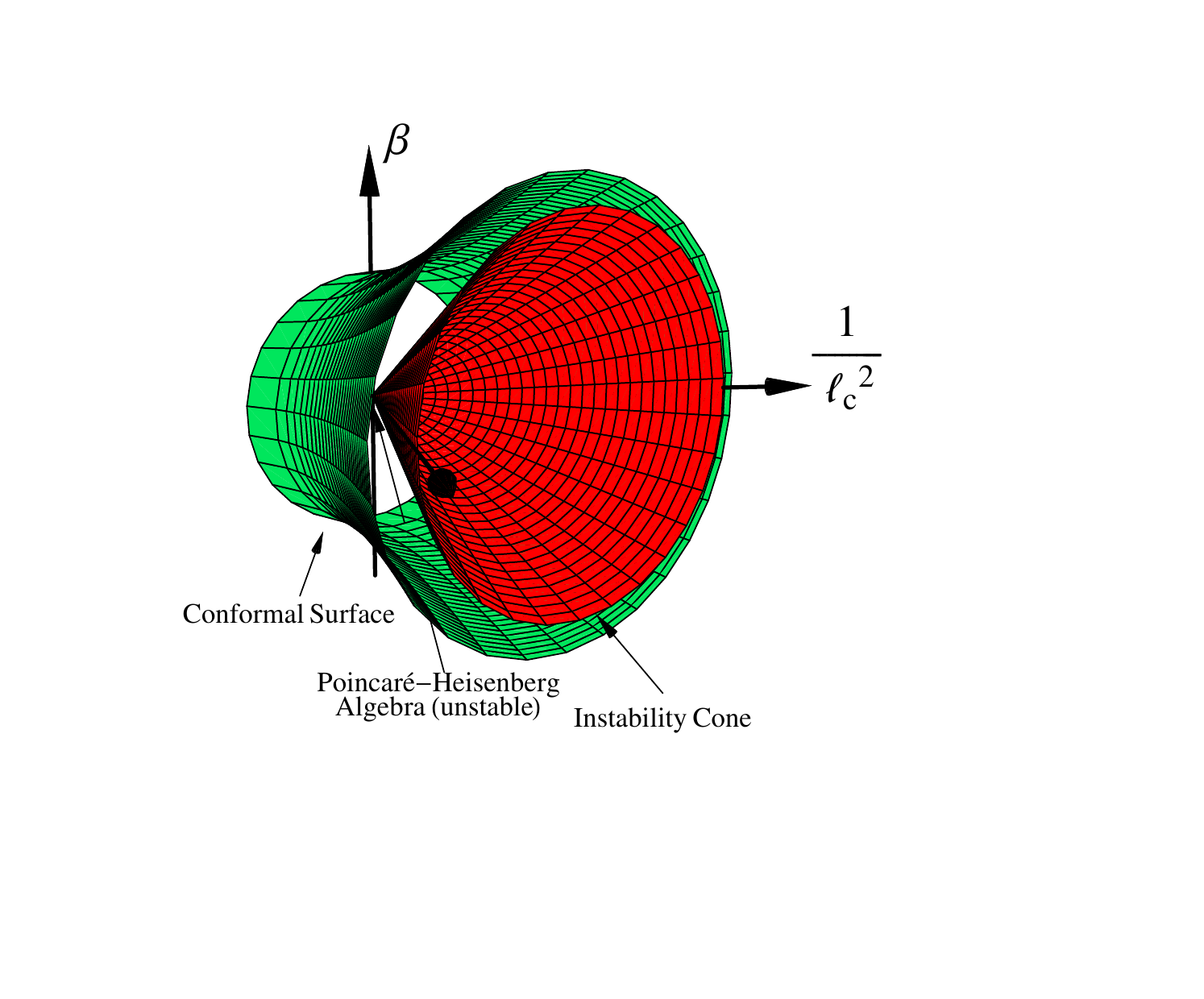}
\caption{The
unmarked arrow is the $\ell_P^2 (=\hbar\alpha_2)$ axis.  
The Poincar\'e-Heisenberg algebra
corresponds to the origin of the parameters space, which coincides with the
apex of the instability cone. In reference to Eq.~(\ref{eq:alpha1}), note
that $\ell_C^2 = \hbar/\alpha_1$. Here, $\beta$ is a dimensionless parameter
that corresponds to a generalisation of the conformal algebra. The SPHA lives
in the entire $(\ell_C,\ell_P,\beta)$ space except for the surface of
instability. The 
SPHA becomes  conformal 
for all values of $(\ell_C,\ell_P,\beta)$ 
that lie on the ``conformal surface''.} 
\end{figure}


\begin{thebibliography}{100}



\bibitem{Isham:1995ci}
C. J. Isham, \textit{Lectures on quantum theory: Mathematical
and structural foundations} (Imperial College Press, Singapore, 
1995).

\bibitem{Ahluwalia:1993dd}
  D.~V.~Ahluwalia,
  ``Quantum measurements, gravitation, and locality,''
  Phys.\ Lett.\  B {\bf 339} (1994) 301
  [arXiv:gr-qc/9308007].


\bibitem{Doplicher:1994zv}
  S.~Doplicher, K.~Fredenhagen and J.~E.~Roberts,
  ``Space-time quantization induced by classical gravity,''
  Phys.\ Lett.\  B {\bf 331} (1994) 39.


\bibitem{Penrose:1996cv}
  R.~Penrose,
  ``On gravity's role in quantum state reduction,''
  Gen.\ Rel.\ Grav.\  {\bf 28} (1996) 581.

\bibitem{Brown:2005}
H.~R.~Brown, Physical relativity: Space-time structure from a dynamical 
perspective (Oxford University Press, Oxford, 2005).

\bibitem{'tHooft:1993gx}
  G.~'t Hooft,
  ``Dimensional reduction in quantum gravity,''
  arXiv:gr-qc/9310026.


\bibitem{Susskind:1994vu}
  L.~Susskind,
  ``The world as a hologram,''
  J.\ Math.\ Phys.\  {\bf 36} (1995) 6377
  [arXiv:hep-th/9409089].

\bibitem{deHaro:2001}
S.~de~Haro, Quantum gravity and the holographic principle
(Universal press - Science publishers, Veenendaal, 2001)


\bibitem{Faddeev:1989}
L.~D.~Faddeev, 
Mathematician's view on the development of physics, Frontiers 
in Physics: High technology and mathematics ed H. A. Cerdeira 
and S. Lundqvist (Singapore: Word Scientific, 1989) pp. 238-46


\bibitem{VilelaMendes:1994zg}
  R.~Vilela Mendes,
  ``Deformations, stable theories and fundamental constants,''
  J.\ Phys.\ A  {\bf 27} (1994) 8091.

\bibitem{Chryssomalakos:2004gk}
  C.~Chryssomalakos and E.~Okon,
  Int.\ J.\ Mod.\ Phys.\  D {\bf 13} (2004) 2003
  [arXiv:hep-th/0410212].


\bibitem{Ahluwalia-Khalilova:2005jn}
  D.~V.~Ahluwalia-Khalilova,
  ``A freely falling frame at the interface of gravitational and quantum
  realms,''
  Class.\ Quant.\ Grav.\  {\bf 22} (2005) 1433
  [arXiv:hep-th/0503141].

\bibitem{Ahluwalia-Khalilova:2005km}
  D.~V.~Ahluwalia-Khalilova,
  ``Minimal spatio-temporal extent of events, neutrinos, and the  cosmological
  constant problem,''
  Int.\ J.\ Mod.\ Phys.\  D {\bf 14} (2005) 2151
  [arXiv:hep-th/0505124].

\bibitem{Gresnigt:2006gq}
  N.~G.~Gresnigt, P.~F.~Renaud and P.~H.~Butler,
  ``The stabilized Poincare-Heisenberg algebra: A Clifford algebra viewpoint,''
  arXiv:hep-th/0611034.

\bibitem{cern:2005}
S. Reucroft and J. Swain,
``Special relativity becomes more general'', 
\textit{CERN Courier}, July/August 2005 p.9.

\bibitem{Collins:2004bp}
  J.~Collins, A.~Perez, D.~Sudarsky, L.~Urrutia and H.~Vucetich,
  ``Lorentz invariance: An additional fine-tuning problem,''
  Phys.\ Rev.\ Lett.\  {\bf 93} (2004) 191301
  [arXiv:gr-qc/0403053].


\bibitem{DiFrancesco:1997ph}
P. Di Francesco, \textit{et al.}, \textit{Conformal field theory} (Springer, New York, 1937).

\bibitem{Garay:1994en}
  L.~J.~Garay,
  ``Quantum gravity and minimum length,''
  Int.\ J.\ Mod.\ Phys.\  A {\bf 10} (1995) 145
  [arXiv:gr-qc/9403008].


\bibitem{Veneziano:1986zf}
  G.~Veneziano,
  ``A stringy nature needs just two constants,''
  Europhys.\ Lett.\  {\bf 2} (1986) 199.


\bibitem{Adler:1999bu}
  R.~J.~Adler and D.~I.~Santiago,
  Mod.\ Phys.\ Lett.\  A {\bf 14} (1999) 1371
  [arXiv:gr-qc/9904026].


\bibitem{deHaroOlle:1997hx}
  S.~de Haro Olle,
  ``Noncommutative black hole algebra and string theory from gravity,''
  Class.\ Quant.\ Grav.\  {\bf 15} (1998) 519
  [arXiv:gr-qc/9707042].


\bibitem{Amelino-Camelia:1997em}
  G.~Amelino-Camelia,
  ``Classicality, matter-antimatter asymmetry, and quantum gravity  deformed
  uncertainty relations,''
  Mod.\ Phys.\ Lett.\  A {\bf 12} (1997) 1387
  [arXiv:gr-qc/9706007].

\bibitem{Maggiore:1993rv}
  M.~Maggiore,
  ``A Generalized uncertainty principle in quantum gravity,''
  Phys.\ Lett.\  B {\bf 304} (1993) 65
  [arXiv:hep-th/9301067].



\bibitem{Sasakura:1999xp}
  N.~Sasakura,
  ``An uncertainty relation of space-time,''
  Prog.\ Theor.\ Phys.\  {\bf 102} (1999) 169
  [arXiv:hep-th/9903146].




\bibitem{Scardigli:1999jh}
  F.~Scardigli,
  ``Generalized uncertainty principle in quantum gravity 
from micro-black  hole
  gedanken experiment,''
  Phys.\ Lett.\  B {\bf 452} (1999) 39
  [arXiv:hep-th/9904025].



\bibitem{Capozziello:1999wx}
  S.~Capozziello, G.~Lambiase and G.~Scarpetta,
  ``Generalized uncertainty principle from quantum geometry,''
  Int.\ J.\ Theor.\ Phys.\  {\bf 39} (2000) 15
  [arXiv:gr-qc/9910017].


\bibitem{Kempf:1994su}
  A.~Kempf, G.~Mangano and R.~B.~Mann,
  Phys.\ Rev.\  D {\bf 52} (1995) 1108
  [arXiv:hep-th/9412167].

\bibitem{Ahluwalia:2000iw}
  D.~V.~Ahluwalia,
  `` Wave particle duality at the Planck scale: Freezing of neutrino 
  oscillations,''
  Phys.\ Lett.\  A {\bf 275} (2000) 31
  [arXiv:gr-qc/0002005].

\bibitem{Buchert:2005xf}
  T.~Buchert,
  ``A cosmic equation of state for the inhomogeneous universe: Can a global
  far-from-equilibrium state explain dark energy?,''
  Class.\ Quant.\ Grav.\  {\bf 22} (2005) L113
  [arXiv:gr-qc/0507028].

\bibitem{Wiltshire:2007jk}
  D.~L.~Wiltshire,
  ``Cosmic clocks, cosmic variance and cosmic averages,''
  arXiv:gr-qc/0702082.

\bibitem{Rasanen:2006kp}
  S.~Rasanen,
  ``Accelerated expansion from structure formation,''
  JCAP {\bf 0611} (2006) 003
  [arXiv:astro-ph/0607626].

\bibitem{Ishibashi:2005sj}
  A.~Ishibashi and R.~M.~Wald,
  ``Can the acceleration of our universe be explained by the effects of
  inhomogeneities?,''
  Class.\ Quant.\ Grav.\  {\bf 23} (2006) 235
  [arXiv:gr-qc/0509108].

\bibitem{Perez:2005gh}
  A.~Perez, H.~Sahlmann and D.~Sudarsky,
  ``On the quantum origin of the seeds of cosmic structure,''
  Class.\ Quant.\ Grav.\  {\bf 23} (2006) 2317
  [arXiv:gr-qc/0508100].


\end{thebibliography}
\end{document}